*Artificial Intelligence and Covid-19*

# Does "AI" stand for augmenting inequality in the era of covid-19 healthcare?

David Leslie,[1] ethics theme lead and ethics fellow, Anjali Mazumder,[1] AI and justice and human rights theme lead, Aidan Peppin,[2] researcher of AI and society, Maria K Wolters,[3] reader in design informatics, Alexa Hagerty,[4] research associate

[1]Alan Turing Institute, London, UK

[2]Ada Lovelace Institute, London, UK

[3]School of Informatics, University of Edinburgh, UK

[4]Centre for the Study of Existential Risk and Leverhulme Centre for the Future of Intelligence, University of Cambridge, Cambridge, UK

Correspondence to: dleslie@turing.ac.uk

*Artificial intelligence can help tackle the covid-19 pandemic, but bias and discrimination in its design and deployment risk exacerbating existing health inequity argue **David Leslie** and **colleagues***

Among the most damaging characteristics of the covid-19 pandemic has been its disproportionate effect on disadvantaged communities. As the outbreak has spread globally, factors such as systemic racism, marginalisation, and structural inequality have created path dependencies that have led to poor health outcomes. These social determinants of infectious disease and vulnerability to disaster have converged to affect already disadvantaged communities with higher levels of economic instability, disease exposure, infection severity, and death. Artificial intelligence (AI) technologies—quantitative models that make statistical inferences from large datasets—are an important part of the health informatics toolkit used to fight contagious disease. AI is well known, however, to be susceptible to algorithmic biases that can entrench and augment existing inequality. Uncritically deploying AI in the fight against covid-19 thus risks amplifying the pandemic's adverse effects on vulnerable groups, exacerbating health inequity.

## Cascading risks and harms

Interacting factors of health inequality include widespread disparities in living and working conditions; differential access to, and quality of, healthcare; systemic racism; and other deep-seated patterns of discrimination. These factors create disproportionate vulnerability to disease for disadvantaged communities, as a result of overcrowding, compelled work, "weathering" (that is, the condition of premature aging and health deterioration due to continual stress), chronic disease, and compromised immune function.[1-3]

This greater vulnerability manifests as increased risks for exposure to covid-19, susceptibility to infection, severity of infection, and death.[4-6] The evidence for these outcomes is rapidly increasing: mortality rates for covid-19 are more than double for those living in more deprived areas[7]; black, Asian, and minority ethnic Britons are up to twice as likely to die if they contract covid-19 in comparison with white Britons.[8 9] When controlling for age, black men and women are more than four times more likely to die than white men and women.[10]

Although AI systems hold promise for improved diagnostic and prognostic decision support, epidemiological monitoring and prediction, and vaccine discovery,[11 12] much research has reported that these systems can discriminate between, and create unequal outcomes in, different sociodemographic groups.[13] The combination of the disproportionate impact of covid-19 on vulnerable communities and the sociotechnical determinants of algorithmic bias and discrimination might deliver a brutal triple punch. Firstly, the use of biased AI models might be disproportionately harmful to vulnerable groups who are not properly represented in training datasets, and who are already subject to widespread health inequality. Secondly, the use of safety critical AI tools for decision assistance in high stakes clinical environments might be more harmful to members of these groups owing to their life and death impacts on them. Lastly, discriminatory AI tools might compound the disproportionate damage inflicted on disadvantaged communities by the SARS-CoV-2 virus.

Despite their promise, AI systems are uniquely positioned to exacerbate health inequalities during the covid-19 pandemic if not responsibly designed and deployed. In this article, we show how the cascading effects of inequality and discrimination manifest in design and use of an AI system (fig 1). To mitigate these effects, we call for inclusive and responsible practices that ensure fair use of medical and public AI systems in times of crisis and normalcy alike.

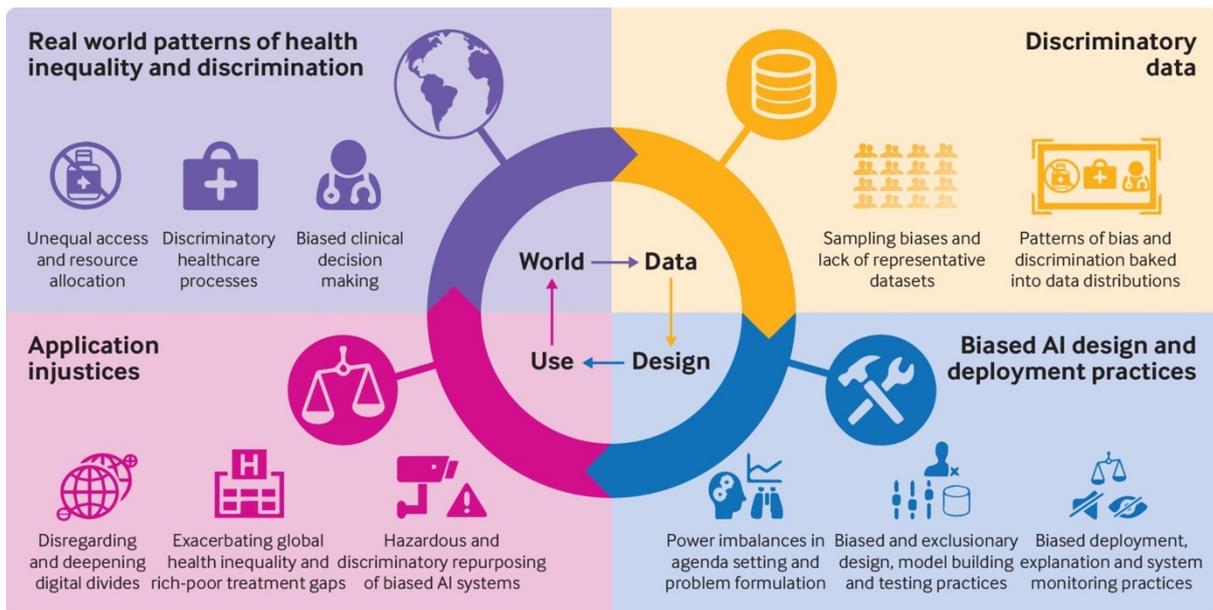

Figure 1: Cascading effects of health inequality and discrimination manifest in the design and use of artificial intelligence (AI) systems

## Embedding inequality in AI systems

Patterns of health inequality permeate AI systems when bias and discrimination become entrenched in the conception, design, and use of these systems across three planes. Discriminatory structures become ingrained in the datasets used to train systems (eg, data from underserved communities are excluded owing to their lack of access to healthcare); deficiencies arise in data representativeness (eg, undersampling of vulnerable populations); and biases crop up across the development and implementation lifecycle (eg, failure to include clinically relevant demographic variables in the model leads to disparate performance for vulnerable subgroups).[14]

## Health discrimination in datasets

AI technologies rely on large datasets. When biases from existing practices and institutional policies and norms affect those datasets, the algorithmic models they generate will reproduce inequities. In clinical and public health settings, biased judgment and decision making, as well as discriminatory healthcare processes, policies, and governance regimens can affect electronic health records, case notes, training curricula, clinical trials, academic studies, and public health monitoring records. During clinical decision making, for example, well established biases against members of marginalised groups, such as African American[15] [41] and LGBT[16] [17] patients, can enter the clinical notes taken by healthcare workers during and after examination or treatment. If these free text notes are then used by natural language processing technologies to pick up symptom profiles or phenotypic characteristics, the real world biases that inform them will be silently tracked as well.

The datasets which are the basis of data driven AI and machine learning models thus reflect complex and historically situated practices, norms, and attitudes. This means that inferences drawn from such medical data by AI models to be used for diagnosis or prognosis might incorporate the biases of previous inequitable practices, and the use of models trained on these datasets could reinforce or amplify discriminatory structures. Risks of this kind of discrimination creep pose special challenges during the covid-19 pandemic. For instance, hospital systems are already using natural language processing technologies to extract diagnostic information from radiology and pathology reports and clinical notes.[42 43 44] As these capacities are shifted onto tasks for identifying clinically significant symptoms of SARS-CoV2 infection,[45] hazards of embedding inequality will also increase. Where human biases are recorded in clinical notes, these discriminatory patterns will probably infiltrate the natural language processing supported AI models that draw on them. Similarly, if such models are also trained using unrepresentative or incomplete data from electronic health records that reflect disparities in healthcare access and quality, the resulting AI systems will probably reflect, repeat, and compound pre-existing structural discrimination.

## Data representativeness

The datasets used to train, test, and validate AI models are too often insufficiently representative of the general public. For instance, datasets composed of electronic health records, genome databases, and biobanks often undersample those who have irregular or limited access to the healthcare system, such as minoritised ethnicities, immigrants, and socioeconomically disadvantaged groups.[18-20] The increased use of digital technologies, like smartphones, for health monitoring (eg, through symptom tracking apps) also creates potential for biased datasets. In the UK, more than 20% of the population aged 15 or older lack essential digital skills and up to 10% of some population subgroups do not own smartphones.[21] Datasets from pervasive sensing, mobile technologies, and social media can under-represent or exclude those without digital access. Whether originating from medical data research facilities or everyday technologies, biased datasets that are linked—such as in biomedical applications that combine pervasive sensing data with electronic health records [22]—will only exacerbate unrepresentativeness.

The prevalence and incidence of diseases and their risk factors often vary by population group. If datasets do not adequately cover populations at particular risk, trained prediction models that are used in clinical AI decision support might have lower sensitivity (true positive rates) for these populations and systematically underdetect the target condition.[23] Every time a prediction model which has been tailored to the members of a dominant group is applied in a "one-size-fits-all" manner to a disadvantaged group, the model might yield suboptimal results and be harmful for disadvantaged people.[24]

The data flows emerging from the covid-19 outbreak present a set of problems that could jeopardise attempts to attain balanced and representative datasets. Tendencies to produce health data silos create a channelling effect where usable electronic health records from patients who have contracted covid-19 overly reflect subpopulations who non-randomly have access to particular hospitals in certain, well-off neighbourhoods. This problem arises because resources needed to ensure satisfactory dataset quality and integrity might be limited to digitally mature hospitals that disproportionately serve a privileged segment of a population to the exclusion of others. Where data from electronic health records resulting from these contexts contribute to the composition of AI training data, problems surrounding discriminatory effects arise. If such dataset imbalances are not dealt with, and if thorough analyses are not performed to determine the limitations of models trained on these data, they will probably not be sufficiently generalisable and transportable. The models will simply underfit members of vulnerable groups whose data were under-represented in the training set, and will perform less well for them.

## Biases in the choices made for AI design and use

Lack of representativeness and patterns of discrimination are not the only sources of bias in AI systems. Legacies of institutional racism and the implicit—often unconscious—biases of AI developers and users might influence choices made in the design and deployment of AI, leading to the integration of discrimination and prejudice into both innovation processes and products.[25]

At the most basic level, the power to undertake health related AI innovation projects is vested with differential privileges and interests that might exacerbate existing health inequities. The sociodemographic composition (that is, class, race, sex, age) of those who set research and innovation agendas often does not reflect that of the communities most affected by the resulting projects.[26 27] This disparity lays the foundation for unequal outcomes from AI innovation. Decisions in setting the agenda include which clinical questions should be reformulated as statistical problems, and which kinds of data centric technologies should be developed. During the covid-19 pandemic this is of particular concern, as the urgency to find solutions and the institutional hierarchies in decision making are at cross purposes with consensus building mechanisms and with the diligence needed to ensure oversight and involvement of the community in setting the agenda.

Once an AI innovation project is underway, choices must be made about how to define target variables and their quantifiable proxies. At this stage of problem formulation, any latent biases of designers, developers, and researchers might allow structural health inequalities and injustices to be introduced in the model via label determinations (that is, choices made in the specification of target variables) that fail to capture underlying complexities of the social contexts of discrimination.[28] This bias was seen in a recent study, which showed that the label choice made by the producers of a commercial insurance risk prediction tool discriminated against millions of African Americans, whose level of chronic illness was systematically mismeasured because healthcare costs were used as a proxy for ill health.[29]

At the stages of extraction, collection, and wrangling of data, measurement errors and faulty data consolidation practices could lead to additional discrimination against disadvantaged communities. For example, if data on skin colour are not collected together with pulse oximetry data, it is almost impossible for AI models to correct for the effect of skin tone on oximetry readings.[46]

Similar discriminatory patterns can pass into design-time processes at the data preprocessing and model construction stages. The decisions made about inclusion of personal data such as age, ethnicity, sex, or socioeconomic status, will affect the way the model performs for vulnerable subgroups. When features such as ethnicity are integrated into models without careful consideration of potential confounders, those models risk identifying as biological, characteristics that have socioeconomic or environmental origins. As a result, structural racism might be integrated into the automated tools that support clinical practice. A well known example is the flawed "race correction" mechanism in commercial spirometer software.[30]

Lastly, AI systems might introduce unequal health outcomes during testing, implementation, and continuing use. For instance, in the implementation phase, clinicians who over-rely on AI decision support systems might take their recommendations at face value, even when these models might be faulty. On the other hand, clinicians who distrust AI decision support systems might discount their recommendations, even if they offer corrections to discrimination. For example, when a decision support model provides pulse oximetry values that have been correctly adjusted for skin tone, the results might conflict with a clinician's own preconceptions about the validity of raw oximetry data. These results might lead the clinician to dismiss the model's recommendation based upon their own potentially biased professional judgment.

## Equity under pressure

During the covid-19 pandemic, demand for rapid response technological interventions might hinder responsible AI design and use.[31] [32] In a living systematic review of over 100 covid-19 prediction models for diagnosis and prognosis, Wynants et al have found that owing to the pressure of rushed research, the proposed systems were at high risk of statistical bias, poorly reported, and overoptimistic. Up to this point, the authors have recommended that none of the models be used in medical practice.[33]

To make matters worse, some hospitals are hurriedly repurposing AI systems (which were developed for use, and trained on data, in situations other than the pandemic) for sensitive tasks like predicting the deterioration of infected patients who might need intensive care or mechanical ventilation.[34] These models run considerable risks of insufficient validation, inconsistent reliability, and poor generalisability due to unrepresentative samples and a mismatch between the population represented in the training data and those who are disparately affected by the outbreak.[35]

AI systems are similarly being swiftly repurposed in non-clinical domains, with tangible consequences for public health. In an attempt to curb the spread of covid-19, the United Sates prison system, for example, has used an algorithmic tool developed for measuring the risk of recidivism to determine which inmates will be released to home confinement. This tool has been shown to exhibit racial biases, and so repurposing it for the management of health risks makes black inmates more likely to remain confined and consequently, subjected to increased exposure to covid-19 infection and disease related death.[36] At the beginning of the second US wave of the pandemic in June, such repurposing took place while the five largest known clusters of covid-19 in the US were at correctional institutions,[37] and against a backdrop of mass incarceration based on historic and systemic racism.[38]

## Conclusion

AI could make a valuable contribution to clinical, research, and public health tools in the fight against covid-19. The widespread sense of urgency to innovate, however, should be tempered by the need to consider existing health inequalities, disproportionate pandemic vulnerability, sociotechnical determinants of algorithmic discrimination, and the serious consequences of clinical and epidemiological AI applications. Without this consideration, patterns of systemic health inequity and bias will enter AI systems dedicated to tackling the pandemic, amplifying inequality, and subjecting disadvantaged communities to increasingly disproportionate harm.

With these dynamics in mind, it is essential to think not just of risks but also remedies (fig 2).

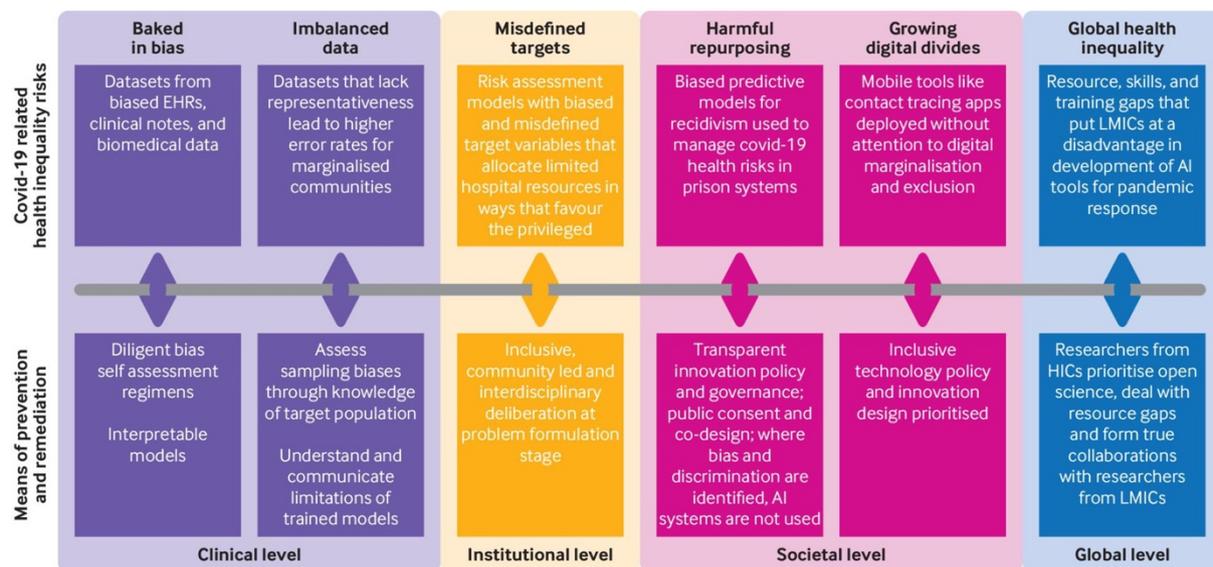

Figure 2: Risks of, and remedies for, developing and deploying artificial intelligence (AI) systems safely. EHRs=electronic health records; HICs=high income countries; LMICs=low and middle income countries

On the latter view, developing and deploying AI systems safely and responsibly in medicine and public health to combat covid-19 requires the following:

- *In technological development:* Incorporation of diligent, deliberate, and end-to-end bias detection and mitigation protocols. Clinical expertise, inclusive community involvement, interdisciplinary knowledge, and ethical reflexivity must be embedded in AI project teams and innovation processes to help identify and remedy any discriminatory factors. Similarly, awareness of the social determinants of disparate vulnerability to covid-19 must be integrated into data gathering practices so that data on socioeconomic status can be combined with other race, ethnicity, and sensitive data to allow for scrutinisation of subgroup differences in processing results.[39 40]
- *In medical and public health practices:* Interpretation of the outputs of AI systems with careful consideration of potential algorithmic biases, and with understanding of the strengths and limitations of statistical reasoning and generalisation. Stakeholders in healthcare must use tools available in public health, epidemiology, evidence based medicine, and applied ethics to evaluate whether specific uses of the quantitative modelling of health data are appropriate, responsible, equitable, and safe.
- *In policy making:* Benefits, limitations, and unintended consequences of AI systems must be considered carefully when setting innovation agendas, without discrimination. Policies will need to be formulated in processes that are open to all stakeholders and prioritise individual and community consent in determining the purpose and path of AI innovation projects.

Finally, as a society, we must deal effectively with systemic racism, wealth disparities, and other structural inequities, which are the root causes of discrimination and health inequalities and evident in algorithmic bias. If we do so, AI can help counter exacerbations of inequalities, instead of contributing to them.

**Key messages**

- The impact of covid-19 has fallen disproportionately on disadvantaged and vulnerable communities, and the use of artificial intelligence (AI) technologies to combat the pandemic risks compounding these inequities
- AI systems can introduce or reflect bias and discrimination in three ways: in patterns of health discrimination that become entrenched in datasets, in data representativeness, and in human choices made during the design, development, and deployment of these systems
- The use of AI threatens to exacerbate the disparate effect of covid-19 on marginalised, under-represented, and vulnerable groups, particularly black, Asian, and other minoritised ethnic people, older populations, and those of lower socioeconomic status
- To mitigate the compounding effects of AI on inequalities associated with covid-19, decision makers, technology developers, and health officials must account for the potential biases and inequities at all stages of the AI process

**Fig 1** Cascading effects of health inequality and discrimination manifest in the design and use of artificial intelligence (AI) systems[

**Fig 2** Risks of, and remedies for, developing and deploying AI systems safely. AI=artificial intelligence; EHRs=electronic health records; HICs=high income countries; LMICs=low and middle income countries.